\documentclass[showpacs,aps,prr,twocolumn,superscriptaddress,amsmath,amssymb,epsfig]{revtex4-2}

\usepackage{graphicx}
\usepackage{epsfig}
\usepackage{color}
\usepackage{dcolumn}% Align table columns on decimal point
\usepackage{bm}

\begin{document}

\title{Failure process of fiber bundles with random misalignment}

 \author{Ferenc Kun}
   \affiliation{Department of Theoretical Physics, Doctoral School of Physics, Faculty of Science and Technology, University of Debrecen, P.O.\ Box 400, H-4002 Debrecen, Hungary}
 \affiliation{HUN-REN Institute for Nuclear Research, PO Box 51, H-4001 Debrecen, Hungary}
  \author{Lynet Allan}
     \affiliation{Department of Theoretical Physics, Doctoral School of Physics, Faculty of Science and Technology, University of Debrecen, P.O.\ Box 400, H-4002 Debrecen, Hungary}
  \author{Attia Batool}
     \affiliation{Department of Physics, School of Natural Sciences, National University of Sciences and Technology, Islamabad 44000, Pakistan}     
    \author{Zsuzsa Danku}
     \affiliation{Department of Theoretical Physics, Doctoral School of Physics, Faculty of Science and Technology, University of Debrecen, P.O.\ Box 400, H-4002 Debrecen, Hungary}
  \author{Gerg\H o P\'al}
     \affiliation{Department of Theoretical Physics, Doctoral School of Physics, Faculty of Science and Technology, University of Debrecen, P.O.\ Box 400, H-4002 Debrecen, Hungary}

 \email{Corresponding author: ferenc.kun@science.unideb.hu}

\begin{abstract}
We investigate the failure process of fiber bundles with structural disorder represented by the random misalignment of fibers. The strength of fibers is assumed to be constant so that misalignment is the only source of disorder, which results in a heterogeneous load distribution over fibers. We show by analytical calculations and computer simulations that increasing the amount of structural disorder a transition occurs from a perfectly brittle behaviour with abrupt global failure to a quasi-brittle phase where failure is preceded by breaking avalanches. The size distribution of avalanches follows a power law functional form with a complex dependence of the exponent on the amount of disorder. In the vicinity of the critical point the avalanche exponent is 3/2, however, with increasing disorder a crossover emerges to a higher exponent 5/2. We show analytically that the mechanical behaviour of the bundle of misaligned fiber with no strength disorder can be mapped to an equal load sharing fiber bundle of perfectly aligned fibers with properly selected strength disorder.
\end{abstract}

\date{\today}

\maketitle

\section{Introduction}
Fibrous materials are prevalent in a wide variety of applications from engineering to biology \cite{fibrous_book_2020,santucci_sub-critical_2004,deschanel_experimental_2009,rosti_statistics_2010,alava_statistical_2006,book_chakrabarti_2015}. Understanding and predicting the mechanical response of these materials is crucial for the design and optimization of numerous products, ranging from composite materials to biological tissues \cite{kawamura_revmodphys_2012,layton_equal_2006,pugno_phenomenological_2008,van_liedekerke_mechanisms_2011,Vas2022}. One of the key tools for studying the mechanical behavior of fibrous materials is the fiber bundle model (FBM) \cite{hansen2015fiber,gleaton2023fiber}, which was introduced by Pierce in 1926 \cite{pierce_1926}. Since then the model has been extended to capture a broad spectrum of material behaviours from plasticity \cite{rinaldi_statistical_2011,raischel_failure_2006} through viscoelasticity \cite{hidalgo_creep_2002,baxevanis_load_2007,danku_PhysRevLett.111.084302,danku_creep_2013,alava_prmat_2020,roy_prres_2019} to thermal effects \cite{roux_thermally_2000,yoshioka_size_2008,hansen_prres_2020} and gained widespread applications to analyze the failure and fracture behavior of heterogeneous materials.
%FBMs provide a simple representation of materials' disorder, hence, it served as an important generic modeling framework to understand the effect of disorder on failure phenomena.
In its simplest form, the model considers a bundle of parallel fibers, all of which are perfectly aligned and subjected to an external load parallel to the fibers' direction. Fibers are assumed to have the same stiffness but a random strength, which is described by a probability distribution. After a fiber fails, its load has to be redistributed over the remaining intact fibers. For load sharing two limiting cases have been widely studied, i.e.\ equal load sharing (ELS) when all fibers have the same load \cite{kloster_burst_1997,hidalgo_avalanche_2009}, and localized load sharing (LLS) when load is redistributed in the local neighborhood of the failed fiber \cite{hansen_burst_1994,newman_time-dependent_2001, biswas_lls_2017,layton_equal_2006,raischel_local_2006,tommasi_localized_2008,hansen_lls_dimension_2015,danku_dim_pre,attia_chaossolit_2022}. 
%The strength distribution of fibers and the load sharing protocol are those elements of the model where material properties can be captured. 
In the basic FBM the strength of fibers is the only source of disorder the amount of which can be controlled by varying e.g.\ the standard deviation of the strength distribution. During the past decades FBMs have provided invaluable insights into the statistical nature of fracture and failure of heterogeneous materials both on the macro- and micro-scales \cite{sornette_elasticity_1989,hansen_burst_1994,kloster_burst_1997,johansen_critical_2000,andersen_tricritical_1997,silveira_comment_1998,kim_phase_2004,kun_extensions_2006,hidalgo_avalanche_2009,menezes-sobrinho_influence_2010,manna_PhysRevE.91.032103,kadar_pre_2017,roy_prres_2019,banerjee_efm_2024}. %Analytic calculations and computer simulations have revealed that under stress controlled loading the failure of FBMs proceeds in bursts of breaking fibers which correspond to crackling events recorded by the acoustic emission technique in laboratory experiments. The probability distribution of 

Despite the utility of the basic fiber bundle model, real-world fibrous materials often exhibit complexities that are not captured by the assumption of perfect alignment. In practice, fibers can be misaligned due to manufacturing processes, environmental factors, or inherent material properties \cite{fibrous_book_2020}. Such misalignment can affect the load distribution among fibers and, consequently, the fracture process of the bundle. To address this, here
we extend the traditional fiber bundle model by incorporating structural disorder in the form of the random misalignment of fibers. 
%This extension allows us to study how the degree of structural disorder influences the macroscopic response and the microscopic process of the failure of the bundle. 
To isolate the effect of structural disorder, in our model misalignment is the only source of disorder, i.e.\ fibers are assumed to have the same strength. By comparing the results of the misaligned fiber bundle model of no strength disorder to the traditional aligned model with strength disorder, we aim to elucidate the impact of fiber orientation variability on the macroscopic properties and failure dynamics of fiber bundles. We show that increasing the degree of misalignment a transition occurs from a perfectly brittle phase, where the first fiber breaking triggers the catastrophic collapse of the bundle, to a quasi-brittle phase, where macroscopic failure is preceded by avalanches of breaking fibers. The brittle to quasi-brittle transition occurs abruptly analogous to first order phase transitions. Computer simulations revealed that in spite of the inhomogeneous load distribution on fibers, the size distribution of avalanches has a power law behaviour with a crossover between two regimes of different exponents. 
%The avalanche size exponents are the same as the mean field exponents of FBMs of aligned fibers. 
For small misalignment and deformation we establish a mapping between FBMs of misaligned fibers of no strength disorder with FBMs of aligned fibers of strength disorder.

%Almost all of the tissues and organs in the human body, such as the bone, skin, tendon, and cartilage, are synthesized and hierarchically organized into fibrous structures with nano-/microsized fibers.

\section{Fiber bundle model with structural disorder}
\begin{figure}
\begin{center}
\epsfig{file=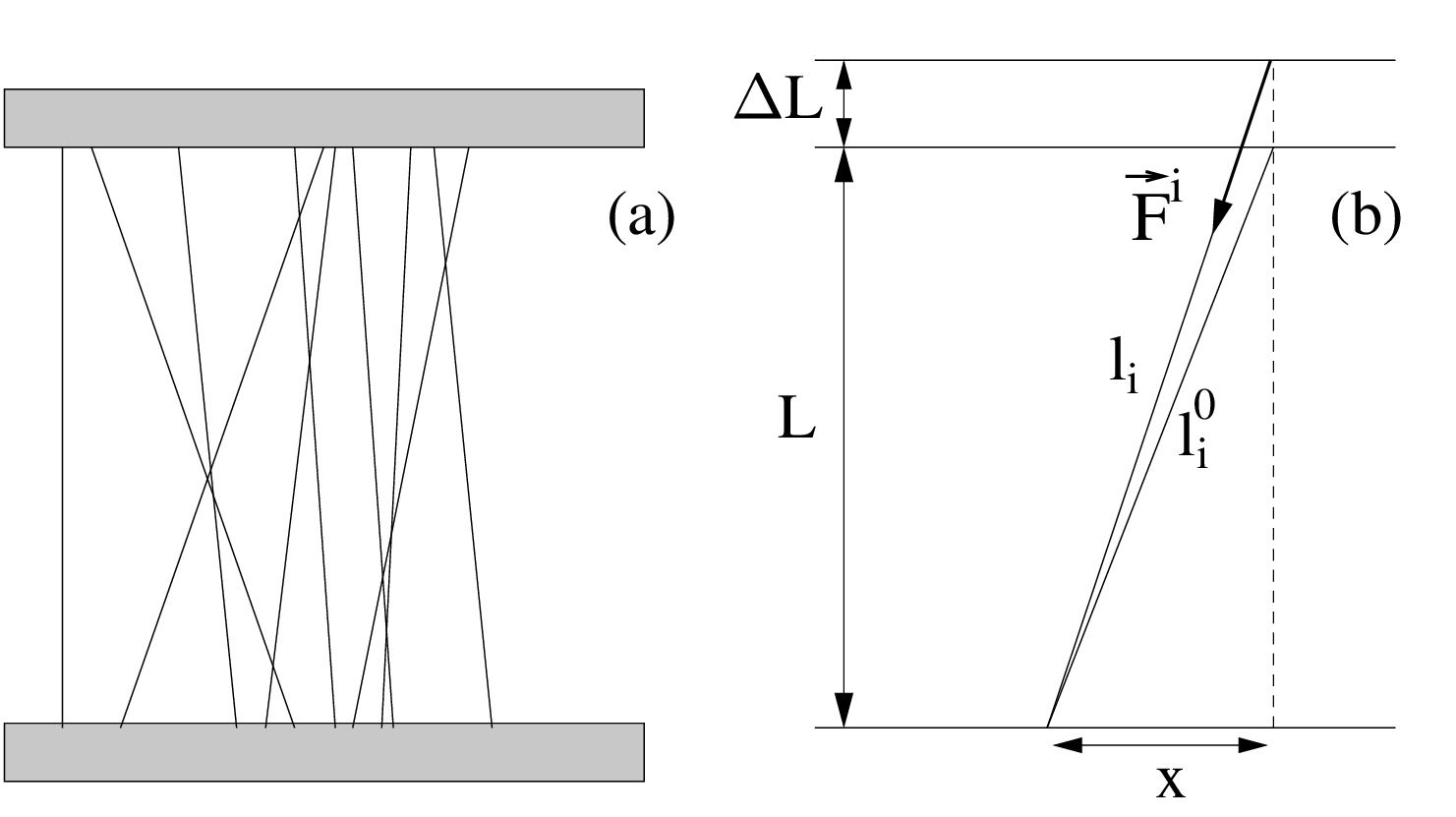,bbllx=0,bblly=0,bburx=700,bbury=400, width=8.5cm}
\caption{Demonstration of the model construction. $(a)$ Bundle of misaligned fibers between two loading plates. External load is applied in the direction perpendicular to the plates. $(b)$ Illustration of the model parameters and variables: initial length $L$ and elongation $\Delta L$ of the bundle, the initial length $l_i^0$ and actual length $l_i$ of a fiber at the global elongation $\Delta L$ of the bundle, the force $\vec{F}^i$ exerted by an elongated fiber. The degree of misalignment of fibers is controlled by the distance $x$ between their two ends along the loading plates.
}   
\label{fig:demo}
\end{center}
\end{figure}

We consider a bundle of $N$ fibers, which are 
assumed to have a perfectly brittle behavior with a stiffness $D$. The bundle is arranged between two parallel loading plates of distance $L$, which defines the initial length of the bundle. The bundle can be loaded in the direction perpendicular to the plates either in a stress or strain controlled way. It is an important feature of the model construction that fibers are not assumed to be parallel to each other, i.e.\ they may have a certain degree of misalignment which is quantified by the distance $x$ of the fibers' two end points along the plates (see Fig.\ \ref{fig:demo} for an illustration). To represent a disordered bundle structure we assume that $x$ is a random variable sampled from a probability density function $p(x)$ over the interval $0\leq x \leq x_{m}$. Note that fibers of the same misalignment $x$ are not necessarily parallel to each other. In the two-dimensional illustration of Fig.\ \ref{fig:demo} tilting in the left and right directions from the bundle axis are equivalent to each other. In three dimensions fibers of a given $x$ value may be rotated to any direction around the bundle axis. Misalignment has the consequence that the initial length $l_i^0$ of fibers becomes also a random variable
\begin{align}
    l_i^0=\sqrt{L^2+x_i^2},
\end{align}
being a monotonically increasing function of $x_i$ ($i=1,\ldots,N$). Figure \ref{fig:demo} illustrates that under a slowly increasing external load the elongation $\Delta l_i$ of fibers also depends on the value of their misalignment $x_i$: at the externally imposed elongation $\Delta L$ of the bundle the fibers get elongated to the length $l_i$, which depends both on $\Delta L$ and $x_i$
\begin{align}
    l_i(\Delta L, x_i) = \sqrt{(L+\Delta L)^2 + x_i^2}.
\end{align}
Consequently, those fibers which are aligned with the load direction $x=0$, suffer the highest elongation $\Delta l_i = l_i-l_i^0 = \Delta L$, however, as $x$ increases $\Delta l_i$ decreases. Fibers are assumed to have a finite load bearing capacity so that they break when their local strain $\varepsilon_i=\Delta l_i/l_i^0$ exceeds a threshold value $\varepsilon_{th}^i$ ($i=1,\ldots, N$). For simplicity, we assume that the misalignment of fibers is the only source of disorder in the bundle so that both the stiffness of fibers $D$ and threshold strain of failure $\varepsilon_{th}$ are assumed to be constant set to the values $D=1$ and $\varepsilon_{th}=0.05$, respectively.

Elongated fibers exert a force $\vec{F}$ which is always parallel to the fibers' actual orientation (see Fig.\ \ref{fig:demo}). To characterize the macroscopic response of the bundle we have to determine how the vertical component, i.e.\ the $y$ component $F_y^i$ of the local force $\vec{F}^i$ of intact fibers varies with increasing elongation $\Delta L$. Taking into account that the angle $\Theta$ a fiber of misalignment $x$ encloses with the $y$ direction evolves with the elongation $\Delta L$ of the bundle as
\begin{align}
\cos{\Theta_i} = \frac{L+\Delta L}{l_i},
\end{align}
the vertical force component of a fiber $i$ can be cast into the form
\begin{align}
    F_y^i = D\left(l_i-l_i^0\right) \frac{L+\Delta L}{l_i}.
    \label{eq:fyfiber}
\end{align}
The total force $F_t$ needed to maintain a globally imposed deformation $\Delta L$ can be obtained as
\begin{align}
    F_t(\Delta L) = \sum_i{}^{\prime} F_y^i = \sum_i{}^{\prime} D\left(l_i-l_i^0\right) \frac{L+\Delta L}{l_i},
    \label{eq:discrete_force}
\end{align}
where the symbol ${}^{\prime}$ indicates that the summation is performed only over those fibers which are intact at the elongation $\Delta L$.

We carried out analytical calculations and computer simulations to understand the macroscopic response and the microscopic failure process of our FBM with random misalignment. To control the amount of structural disorder we considered a uniform distribution $p(x)$ of misalignment $x$  between 0 and $x_m$ with the probability density function
\begin{align}
    p(x) = \frac{1}{x_m}.
    \label{eq:disorder}
\end{align}
The degree of structural disorder due to misalignment of fibers is characterized by the dimensionless ratio $x_m/L$, which was varied in the range $0\leq x_m/L \leq 100$. In the following we demonstrate by analytical and numerical calculations that our FBM with structural disorder but no strength disorder exhibits a complex response both on the macro- and micro-scales. Simulations were performed with $N=10^5$ fibers under stress controlled conditions averaging over $K=1000$ samples at each parameter set.

\section{Macroscopic constitutive behaviour of the bundle}
Assuming strain controlled loading the generic form of the macroscopic constitutive equation of the bundle can be given in a closed analytical form. It follows from the geometrical arrangement of fibers that at an externally imposed elongation of the bundle $\Delta L$, the local elongation of single fibers $\Delta l_i(\Delta L, x_i)$ ($i=1,\ldots , N$) is a decreasing function of their misalignment $x_i$: fibers which are aligned with the load direction $x=0$ have the highest elongation $\Delta l_i = \Delta L$, and $\Delta l_i$ monotonically decreases with increasing $x_i$. Since the failure strain $\varepsilon_{th}$ is assumed to be constant, it follows that fibers break in the increasing order of their misalignment. Consequently, at an elongation $\Delta L$ only those fibers contribute to the total force $F_t$ which have a misalignment $x$ larger than a threshold value $x_l(\Delta L)$, which depends on $\Delta L$. These contributions can be summed up by an integral
\begin{align}
    F_t(\Delta L) = N\int_{x_l(\Delta L)}^{x_m}F_y(\Delta L,x)p(x)dx,
    \label{eq:constit_int}
\end{align}
which yields the force-elongation relation $F_t(\Delta L)$ of the bundle. At the beginning of the loading process the global strain of the bundle $\Delta L/L$ falls below the failure threshold  $\Delta L/L<\varepsilon_{th}$, that's why no fiber breaking can occur. Consequently, in this regime the lower bound of the integral is $x_l=0$. When $\Delta L$ exceeds $\Delta L_{min}=L\varepsilon_{th}$, all fibers break which have a misalignment below 
\begin{align}
    x_c = \sqrt{\frac{(L+\Delta L)^2-AL^2}{A-1}},
\end{align}
where $A=(1+\varepsilon_{th})^2$. Hence, the lower bound $x_l$ of the integral form of the constitutive equation Eq.\ (\ref{eq:constit_int}) can be cast into the general form
\begin{align}
    x_l(\Delta L) = 
    \left\{
\begin{array}{ccc}
   0   & \mbox{if} & \Delta L\leq L\varepsilon_{th}, \\
   x_c & \mbox{if} & \Delta L> L\varepsilon_{th}.
\end{array}
    \right.
\end{align}
\begin{figure}
\begin{center}
\epsfig{file=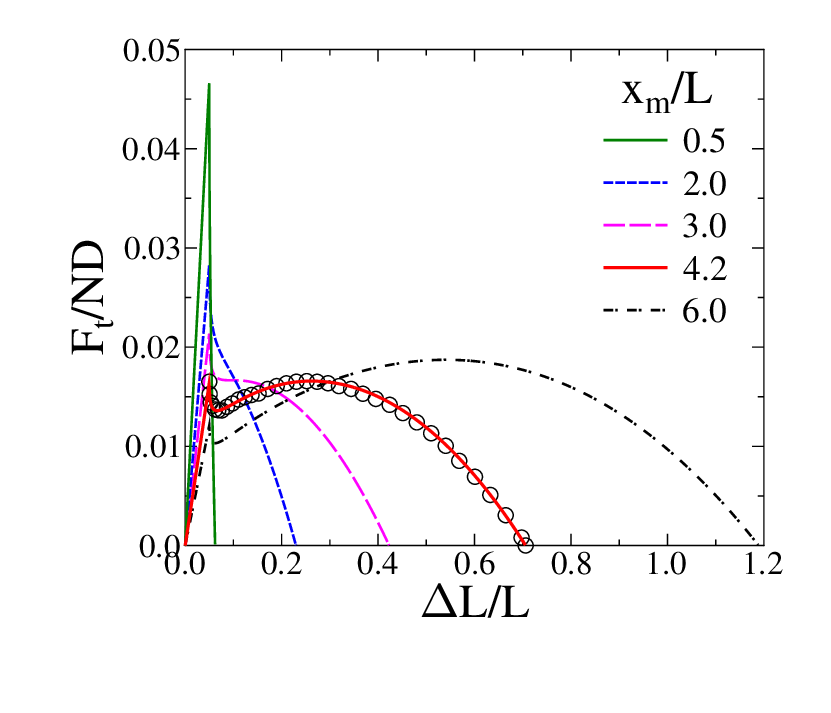,bbllx=30,bblly=30,bburx=390,bbury=330, width=8.6cm}
\caption{Force-elongation diagrams of the model at several values of the disorder parameter $x_m$. As $x_m$ increases the initial sharp peak gets gradually reduced and a second local maximum develops. At the $x_m$ value where the second maximum becomes higher than the first one, a transition occurs from perfectly brittle to quasi-brittle behaviour. For $x_m/L=4.2$ computer simulations (represented by the circles) are compared to the analytical results. An excellent agreement can be observed.}   
\label{fig:constit}
\end{center}
\end{figure}
For the case of uniformly distributed misalignment, explicit results can be obtained by substituting the expressions of $F_y(\Delta L, x)$ and $p(x)$ from Eqs.\ (\ref{eq:fyfiber},\ref{eq:disorder}) into Eq.\ (\ref{eq:constit_int})
\begin{align}
    F_t(\Delta L) = \frac{ND(L+\Delta L)}{x_m} \int_{x_l}^{x_m} \left[1-\frac{\sqrt{L^2+x^2}}{\sqrt{(L+\Delta L)^2+x^2}}\right]dx.
    \label{eq:constit}
\end{align}
Figure \ref{fig:constit} illustrates the force-elongation diagram of the bundle, where the integral of Eq.\ (\ref{eq:constit}) was calculated numerically at each elongation $\Delta L$ for different amounts of structural disorder $x_m$. It can be seen that for small deformations the response of the system is linear, however, the effective stiffness of the bundle, i.e.\ the slope of the initial straight line, depends on the amount of disorder $x_m$.
Since all fibers have the same strength, in the limit of low structural disorder the $F_t(\Delta L)$ curves have a sharp maximum followed by a sudden drop of the force, which implies that the system would abruptly collapse in a stress controlled experiment when reaching the maximum of $F_t$. The decreasing regime of the constitutive curves can only be realized in strain controlled experiments. The largest value $\Delta L_{max}$ of $\Delta L$, where $F_t$ reaches zero, is determined by the upper bound of the misalignment $x_m$ in the form
\begin{align}
    \Delta L_{max} = -L+\sqrt{L^ 2+(A-1)(L^2+x_m^2)}.
\end{align}
With increasing disorder $x_m$ the initial brittle peak becomes lower and the maximum elongation $\Delta L_{max}$ of complete distraction increases, while in between a second local maximum of $F_t(\Delta L)$ develops. Further increasing $x_m$ the second maximum becomes eventually dominating, i.e.\ it becomes the global maximum of the force-elongation curve while the brittle peak completely diminishes.

The uniform distribution of misalignment values has the advantage that in the limit of small deformation $\Delta L\ll L$ the constitutive curve can be analyzed analytically. Taylor expansion of the denominator of the integrand in Eq.\ (\ref{eq:constit}) yields the approximate expression
\begin{align}
    \sqrt{(L+\Delta L)^2+x^2} \approx \sqrt{L^2+x^2}\left(1+\frac{L\Delta L}{L^2+x^2}\right).
\end{align}
Substituting it into Eq.\ (\ref{eq:constit}) the integral form of the constitutive equation simplifies to 
\begin{align}
    F_t(\Delta L) \approx \frac{ND(L+\Delta L)}{x_m}\int_{x_l}^{x_m} \frac{L\Delta L}{L^2+x^2}dx,
\end{align}
from which we obtain the closed form
\begin{align}
    F_t(\Delta L) \approx \frac{NDL}{x_m} \arctan{\left(\frac{x_m}{L}\right)}\Delta L.
\end{align}
It can be observed that for small deformations $\Delta L\ll L$, the constitutive behaviour of the system is linear in $\Delta L$, however, its effective stiffness $Y_{eff}$, i.e.\ the slope of the initial straight line, depends on the degree of disorder
\begin{align}
    Y_{eff} = \frac{NDL}{x_m} \arctan{\left(\frac{x_m}{L}\right)}.
\end{align}
Using the Taylor series of the $\arctan$ function as $\arctan{x}\approx x-x^3/3$ the effective stiffness can be cast into the form
\begin{align}
    Y_{eff} = ND\left(1-\frac{1}{3}\frac{x_m^2}{L^2}\right).
    \label{eq:stiffness}
\end{align}
This result shows that in the limit of zero structural disorder $x_m\to 0$, the bundle stiffness is equal to that of the fibers. As $x_m$ increases more-and-more fibers have a high angle with the load direction which gradually reduces the stiffness of the system. 

\begin{figure}
\begin{center}
\epsfig{file=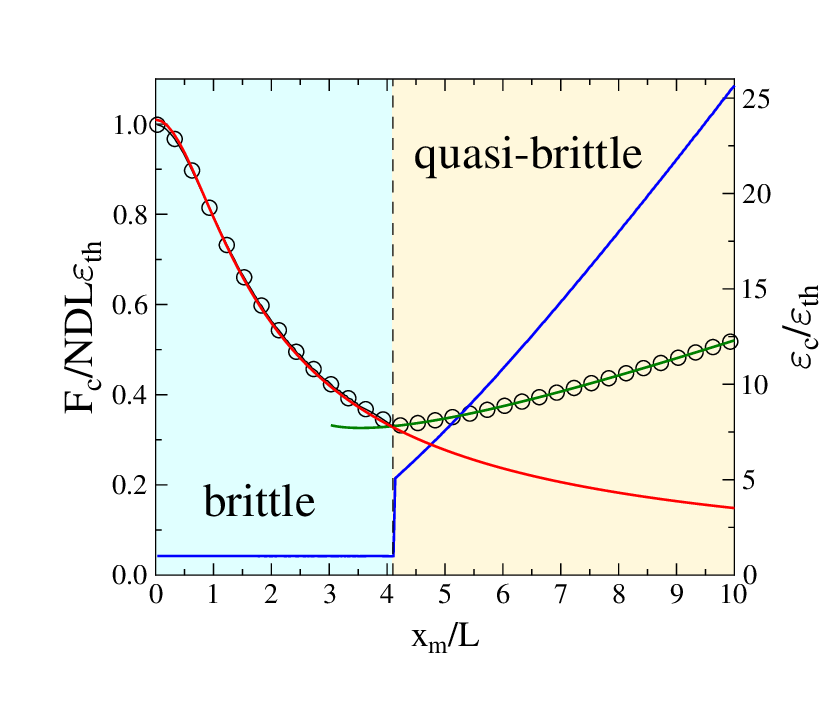,bbllx=25,bblly=25,bburx=395,bbury=310, width=8.6cm}
\caption{The critical force $F_c$ (circles, left vertical axis) and critical strain $\varepsilon_c=\Delta L_c/L$ (blue line, right vertical axis) at which failure occurs under stress controlled loading as function of the amount of structural disorder $x_m/L$. The local minimum of $F_c$ marks the point $x_c$, which separates the phases of perfectly brittle and quasi-brittle failure. The red line represents the analytic results Eq.\ (\ref{eq:fcrit}) for the critical force in the perfectly brittle regime. An excellent agreement of the numerical and analytical results can be observed. The green line indicates the value of the second maximum of the force-elongation curve, starting at about $x_m/L\approx 3.0$, when it first occurs. Above the critical disorder $x_c$, this maximum coincides with $F_c$ (circles).
\label{fig:compare_fcrit}}
\end{center}
\end{figure}

At sufficiently low disorder the initial peak is the global maximum of the $F_t(\Delta L)$ curve, where immediate abrupt failure occurs in a stress controlled experiment. In this disorder regime the overall response of the system is perfectly brittle. However, when $x_m$ exceeds a threshold value $x_c=4.095\pm 0.005$ the second maximum becomes higher than the first brittle peak, which implies that after the onset of fiber breaking the bundle gets stabilized. As the load is further increased, global failure of the bundle is approached through the gradual accumulation of damage, making the bundle quasi-brittle. Our results show that varying the amount of structural disorder, in our FBM a brittle - quasi-brittle transition emerges at the critical disorder $x_c$. To obtain a quantitative characterization of the disorder driven transition, we determined numerically the position $\Delta L_c$ and the value $F_c$ of the global maximum of the force-elongation curve as a function of $x_m$. It can be observed in Fig.\ \ref{fig:compare_fcrit} that first $F_c$ decreases with increasing $x_m$ and reaches a minimum at a specific value of the disorder $x_c$, followed by an increasing branch of the curve. The minimum and the increasing regime of $F_c$ indicate that the second maximum of the $F_t(\Delta L)$ curve overcomes the first one and it controls the macroscopic failure of the bundle. At the same time the critical elongation $\Delta L_c$ corresponding to the maximum force $F_c$, remains constant in the brittle phase, then it makes a finite jump to the second maximum at $x_c$, and finally increases with increasing disorder. Hence, the point $x_c$ can be identified as the critical disorder where the transition from the perfectly brittle to the quasi-brittle behaviour occurs. In Figure \ref{fig:compare_fcrit} the value of the second maximum of the force-elongation curve is also highlighted which first occurs at $x_m/L\approx 3.0$ and coincides with the value of $F_c$ above the critical disorder $x_c$.

The maximum of the force $F_c$ determines the fracture strength of the bundle, which can simply be obtained in the brittle regime as
\begin{align}
    F_c = Y_{eff} L\varepsilon_{th},
    \label{eq:fcrit}
\end{align}
where $Y_{eff}$ has to be substituted from Eq.\ (\ref{eq:stiffness}).
Figure \ref{fig:compare_fcrit} presents a comparison of the numerical value of the peak of $F_t(\Delta L)$ with the closed analytical expression Eq.\ (\ref{eq:fcrit}) of $F_c$ as a function of the disorder parameter $x_m$. An excellent agreement can be observed in the brittle phase $x_m\leq x_c$. 
 
\section{Avalanches of breaking fibers}
%The integral form of the constitutive equation corresponds to the limit of infinite bundle size. For a finite bundle of $N$ fibers with random misalignment values $x_i$ ($i=1,\ldots, N$) the discrete form of the force-elongation expression is provided by Eq.\ (\ref{eq:discrete_force}) 

Under stress controlled loading the failure process of heterogeneous materials is accompanied by bursts of local breaking events, which is well captured by the classical FBM. It has been shown that in FBMs, composed of perfectly aligned fibers and loaded quasi-statically parallel to the fibers' direction, fibers break in avalanches which are analogous to the acoustic outbreaks registered in laboratory measurements \cite{kloster_burst_1997,pradhan_crossover_2005-1,pradhan_crossover_2006,hidalgo_avalanche_2009,biswas_lls_2017,book_chakrabarti_2015}. Analytical calculations have revealed that under equal load sharing conditions the size distribution of avalanches has a power law functional form where the exponent exhibits a high degree of universality, i.e.\ the value of the exponent is $\tau=5/2$ for a broad class of disorder distributions \cite{kloster_burst_1997,pradhan_crossover_2005-1,hidalgo_avalanche_2009}. When the load sharing is localized to nearest neighbor fibers a strong load concentration occurs which leads to a more brittle response where the bundle can tolerate only small sized avalanches with a rapidly decaying distribution which also depends on the probability distribution of the strength of fibers \cite{newman_time-dependent_2001,biswas_lls_2017,layton_equal_2006,raischel_local_2006,tommasi_localized_2008,hansen_lls_dimension_2015,danku_dim_pre,attia_chaossolit_2022}.

Here we want to understand how the disordered structure of the bundle affects the statistics of failure avalanches under a quasi-statically increasing external load. It is a very important consequence of the misalignment of fibers that even if the bundle is loaded between stiff plates the load sharing is not equal in the sense that after the breaking of a fiber the load increment an intact fiber receives depends on its misalignment $x$. The rigidity of the loading plates ensures that the load redistribution is global, i.e.\ each intact fiber shares the increased load, however, not equally. To overcome this computational difficulty, when exploring avalanches of fiber breakings, we worked out the following simulation technique:
\begin{itemize}
    \item Initially, random misalignment values $x_i$ ($i=1,\ldots, N$) are generated according to the probability distribution $p(x)$.
    \item The critical elongation  $\Delta L_i^c$ ($i=1,\ldots,N$) at which fibers break is calculated from the condition
    \begin{align}
        \frac{l_i(\Delta L_i^c, x_i)-l_i^0(x_i)}{l_i^0(x_i)} = \varepsilon_{th},
    \end{align}
    which yields
    \begin{align}
            \Delta L_i^c = -L +\sqrt{L^2+(A-1)(L^2+x_i^2)}.
            \label{eq:delta_l}
    \end{align}
    \item Fibers are sorted in the increasing order of their critical elongation $\Delta L_i^c$. Since $\Delta L_i^c$ is a monotonous function of the corresponding misalignment $x_i$ of fibers, sorting can also be done already for the $x_i$ values, which is then followed by the calculation of the critical elongation.
\begin{figure}
\begin{center}
\epsfig{file=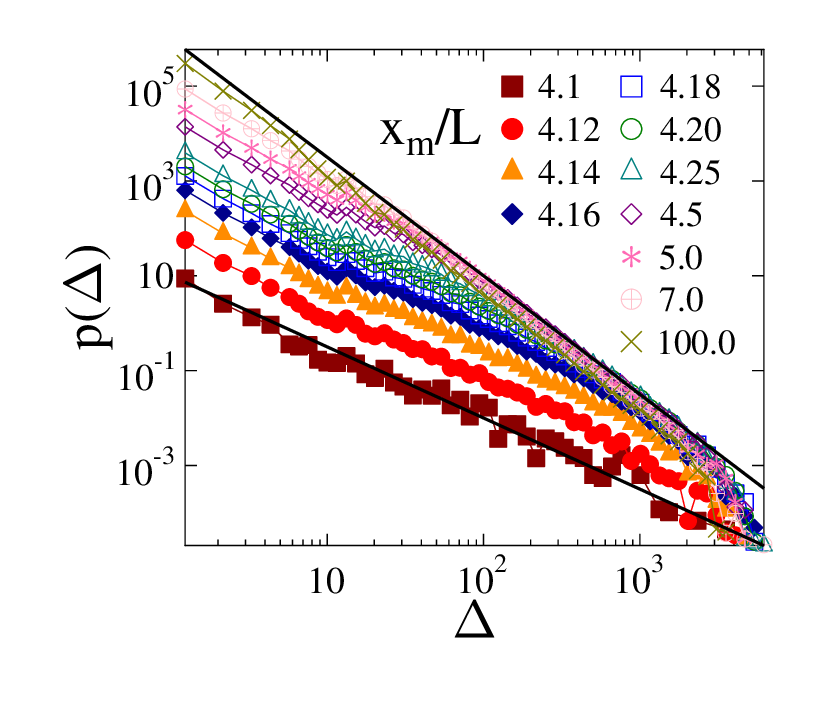,bbllx=35,bblly=35,bburx=380,bbury=330, width=8.5cm}
\caption{Avalanche size distributions $p(\Delta)$ obtained at different values of $x_m$ covering a broad range from the vicinity of the critical point $x_c$ up to $x_m/L=100$. The straight lines represent power laws of exponents $\tau=3/2$ and $\tau=5/2$. 
}   
\label{fig:avaldist}
\end{center}
\end{figure}

    \item Calculation of the discrete force-elongation curve of the bundle at the sorted critical elongations of fibers
    \begin{align}
        F_t(\Delta L_i^c) =\sum_{j=i}^N F_y^j(\Delta L_i^c).
    \end{align}
    Here the summation index $j$ on the right hand side starts from $i$ since all the other fibers of the sorted sequence of $\Delta L^c_j$ values have lower critical elongation $\Delta L_j^c <\Delta L_i^c$ for $j<i$, and hence, are broken at $\Delta L_i^c$, where the force kept by the intact fibers is calculated.
    \item The curve of $F_t(\Delta L_i^c)$ has strong fluctuations where the decreasing regimes can only be realized in strain controlled experiments. Under stress controlled conditions, the load can be increased up to the next local maximum of $F_t(\Delta L_i^c)$. When this fiber is removed those fibers which are in the subsequent valley of the fluctuating curve, cannot keep the increased load, and hence, break. To uncover avalanches of triggered breakings, after the external load increased to a local maximum of the force-elongation curve, we search for the next maximum which is greater than the current one and remove the fibers of the sorted sequence in between as an avalanche of breakings. The size of the avalanche $\Delta$ is determined by the number of fibers breaking in the triggered sequence. 
\end{itemize}
When the global maximum of the force-elongation curve is reached during the loading process, a catastrophic avalanche is initiated which wipes away all the remaining intact fibers. Until the first peak is the highest point of the $F_t(\Delta L)$ curve, this catastrophic avalanche is triggered already by the first fiber breaking, which indicates the brittle character of failure. However, as the second maximum becomes higher than the first one, the evolution of the fracture process drastically changes. The first avalanche, generated between the two maxima, is still large, i.e.\ it comprises a macroscopic fraction of fibers, however, it stops and the bundle gets stabilized, and retains part of its load bearing capacity. Further increasing the external load the catastrophic avalanche is approached through a sequence of stable avalanches. 

For testing purposes we determined the constitutive curve by computer simulations recording the load $F_t(\Delta L_i^c)$ of the bundle at each critical elongation $\Delta L_i^c$ ($i=1,\ldots, N$), where the fibers break. In Fig.\ \ref{fig:constit} the numerically obtained $F_t(\Delta L_i^c)$ curve is compared to the analytical result for $x_m=4.2$, where an excellent agreement is obtained. 

To characterize the statistics of the size $\Delta$ of breaking avalanches we determined the size distribution $p(\Delta)$ for several values of the maximum misalignment $x_m$ covering a broad range from the vicinity of the critical point $x_c$ up to $x_m/L=100$. It can be seen in Fig.\ \ref{fig:avaldist} that $p(\Delta)$ exhibits a power law behaviour 
\begin{align}
    p(\Delta) \sim \Delta^{-\tau},
\end{align}
where the value of the exponent $\tau$ has a complex dependence on the degree of disorder $x_m$. In the close vicinity of the critical disorder $x_c$ the distribution $p(\Delta)$ is composed of a single power law of a relatively low exponent $\tau=3/2$. The low value of $\tau$ indicates the high frequency of large avalanches. Comparing to the behaviour of the constitutive curve $F_t(\Delta L)$ in Fig.\ \ref{fig:constit}, these avalanches are generated around the second maximum of the constitutive curve where the system has a high susceptibility to load increments. Increasing $x_m$ the system becomes more and more quasi-brittle, i.e.\ the avalanches are generated along a longer and longer non-linear regime of $F_t(\Delta L)$ before its maximum. It has the consequence that the avalanche size distribution undergoes a crossover, i.e.\ for small avalanches the low exponent $\tau=3/2$ prevails, however, for the large ones a second power law regime emerges with a higher exponent $\tau=5/2$. The avalanche size $\Delta_c$ corresponding to the crossover point can be determined as the intersection point of the two power laws fitted to the numerical curves (see Fig.\ \ref{fig:avaldist}). Further increasing the amount of disorder $x_m$, the regime of the lower exponent shrinks, i.e.\ the crossover avalanche size $\Delta_c$ decreases, and the distribution becomes again a single power law of exponent $\tau=5/2$ sufficiently far from the critical point. 

\begin{figure}
\begin{center}
\epsfig{file=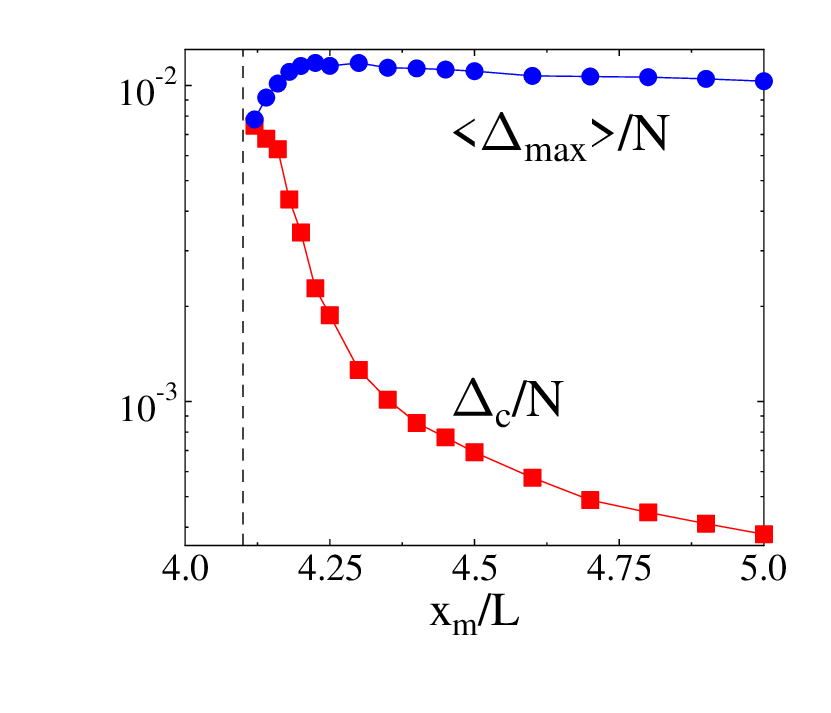,bbllx=40,bblly=30,bburx=380,bbury=330, width=8.5cm}
\caption{The avalanche size $\Delta_c$ corresponding to the crossover point and the average size of the largest avalanche $\left<\Delta_{max}\right>$ as function of $x_m/L$. The vertical dashed line indicates the value of the critical disorder $x_c$.
}   
\label{fig:crossover}
\end{center}
\end{figure} 
In order to give a quantitative characterization of the evolution of the crossover of the burst size distribution with the degree of disorder we determined the avalanche size $\Delta_c$ corresponding to the crossover point. Power laws were fitted to the regime of small and large avalanches and the value of $\Delta_c$ was obtained as the position of the intersection point of the two straight lines on the double logarithmic plot (see Fig.\ \ref{fig:avaldist}).
For this purpose, first the ranges of avalanche sizes with different exponents were  estimated and then the power law fitting was carried out using the maximum likelihood method \cite{newman_siam_powlow}, finally, the crossover avalanche size $\Delta_c$ was determined as the position of the intersection point of the two power laws. Figure \ref{fig:crossover} demonstrates that as $x_m$ approaches $x_c$ from above $\Delta_c$ increases and coincides with the average size of the largest avalanche $\left<\Delta_{max}\right>$, which means that the entire distribution is described by a power law of exponent $\tau=3/2$. Increasing $x_m$ the value of $\Delta_c$ gradually decreases.

\begin{figure}
\begin{center}
\epsfig{file=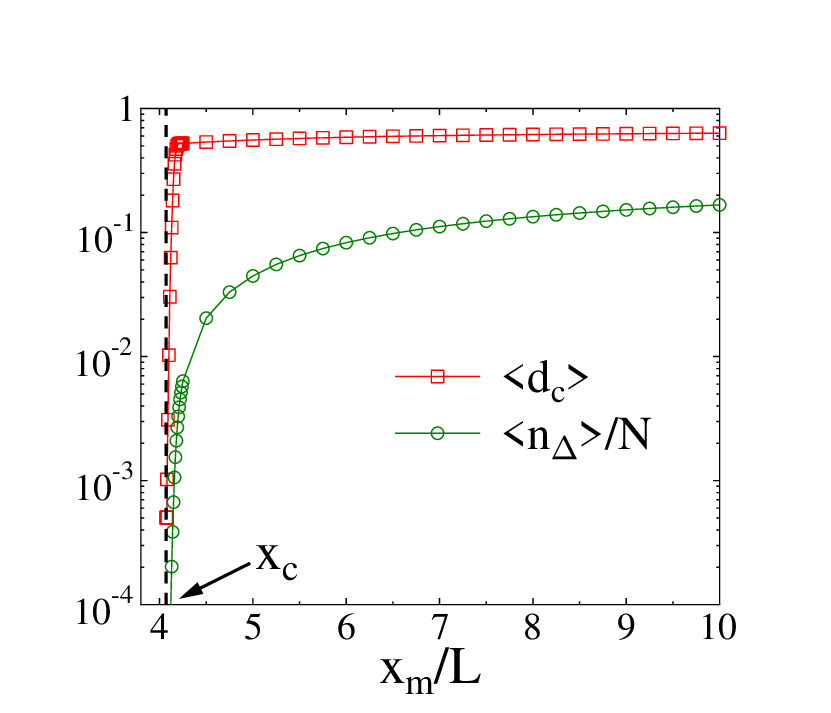,bbllx=20,bblly=0,bburx=360,bbury=300, width=8.5cm}
\caption{Average damage $\left<d_c\right>$ accumulated up to failure and the average number of avalanches $\left<n_{\Delta}\right>$ as function of the degree of disorder $x_m/L$. 
%Inset: The average size of the first avalanche $\left<\Delta_1\right>$ and of the catastrophic one $\left<\Delta_c\right>$ as function of $x_m$. Avalanche sizes are scaled with the number of fibers $N$ of the bundle.
}   
\label{fig:aver_aval}
\end{center}
\end{figure}

To follow how the overall behaviour of the system evolves with increasing structural disorder, we determined the number of avalanches $n_{\Delta}$ that occurred up to failure (excluding the catastrophic avalanche) and the total damage $d_c$ they caused. The amount of damage is obtained as the fraction of broken fibers, i.e.\ the sum of avalanche sizes that occurred up to failure divided by the initial number of fibers 
\begin{align}
    d_c = \frac{1}{N}\sum_i \Delta_i,
\end{align}
which falls between 0 and 1. Figure \ref{fig:aver_aval} shows that for $x_m<x_c$ no damage can occur, hence, $d_c=0$ holds. Above the critical point $x_m>x_c$ the system can support a growing amount of damage, hence, the average damage $\left<d_c\right>$ rapidly increases and saturates at around $\left<d_c\right>\approx 0.66$ for large $x_m$. Damage is formed by the consecutive avalanches of breaking events, hence, the average number of avalanches $\left<n_{\Delta}\right>$ exhibits the same qualitative behaviour as $\left<d_c\right>$, i.e.\ it has a zero value $\left<n_{\Delta}\right>=0$ below the critical disorder $x_c$, while it increases monotonically above $x_c$. We carefully analyzed the functional form of both $\left<d_c\right>$ and  $\left<n_{\Delta}\right>$ as a function of the distance from the 
critical point $x_m-x_c$ in the regime $x_m>x_c$, however, neither power law nor exponential dependence could be pointed out. The results demonstrate that in our system the brittle to quasi-brittle transition does not show an analogy to continuous phase transitions in the sense that no scaling emerges in terms of the distance from the critical point. Both characteristic quantities $\left<d_c\right>$ and $\left<n_{\Delta}\right>$, and the critical elongation of failure $\Delta L_c$ have a finite jump at the critical disorder indicating a first order type transition from the brittle to the quasi-brittle phase of the system.

\section{Relation to FBMs with strength disorder}
\begin{figure}
\begin{center}
\epsfig{file=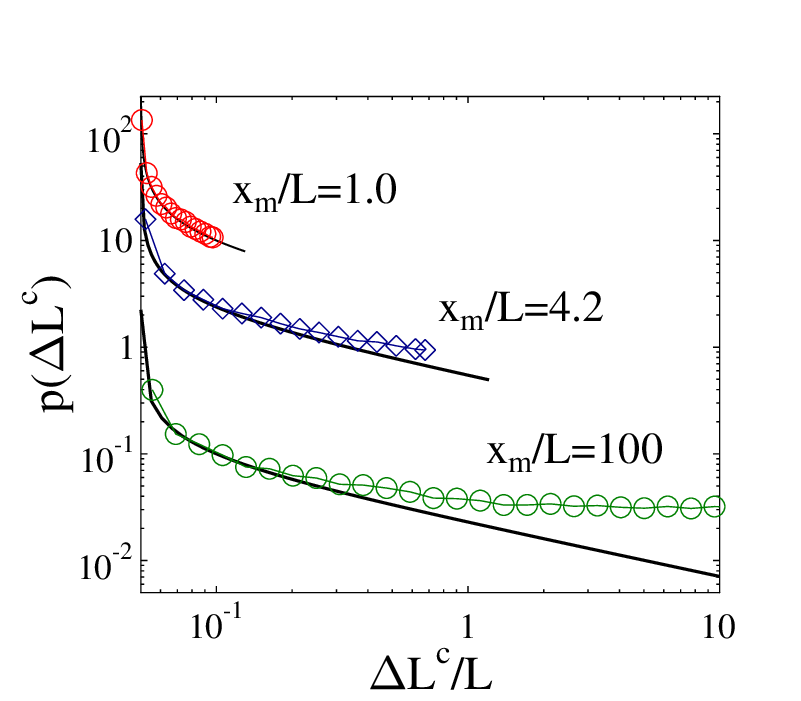,bbllx=0,bblly=0,bburx=360,bbury=310, width=8.5cm}
\caption{Probability distribution $p(\Delta L^c)$ of the critical elongation $\Delta L^c$ of fibers at three different values of the upper bound of the misalignment $x_m/L$. The bold black lines represent curves of the analytical expression Eq.\ (\ref{eq:dist_dlc}) with the corresponding parameters.
\label{fig:dist_dl}}
\end{center}
\end{figure}

A special feature of our fiber bundle model is that it does not contain strength disorder, i.e.\ the fibers have the same threshold strain $\varepsilon_{th}$ at which they fail irreversibly. However, the misalignment of fibers introduces structural disorder which in turn results in avalanches of triggered breaking events. At a sufficiently high structural disorder the system can stabilize after avalanches so that global failure is approached through a random sequence of breaking bursts and its overall strength increases with the amount of disorder. To understand how the increasing structural disorder gives rise to the emergence of the brittle - quasi-brittle transition, we analyzed how the failure elongation $\Delta L^c$ and force $F_y$ kept by the fibers depend on the degree of their misalignment $x$. Starting from Eq.\ (\ref{eq:delta_l}) one can show that for $\varepsilon_{th}\ll 1$ and $x/L\ll 1$ the critical elongation $\Delta L^c$ of breaking takes the simple form
\begin{align}
    \Delta L^c \approx L\varepsilon_{th}\left(1+\frac{x^2}{L^2}\right).
\end{align}
The random misalignment $x$ of fibers introduces randomness for their critical elongation $\Delta L^c$. Since $x$ is uniformly distributed between $0$ and $x_m$, the probability density function of $\Delta L^c$ can be cast into the form
\begin{align}
    p(\Delta L^c) = \frac{1}{2x_m\varepsilon_{th}}\left(\frac{\Delta L^c}{L\varepsilon_{th}}-1\right)^{-1/2},
    \label{eq:dist_dlc}
\end{align}
where $\Delta L^c$ spans the interval from $L\varepsilon_{th}$ and $L\varepsilon_{th}(1+x_m^2/L^2)$. It can be seen in Fig.\ \ref{fig:dist_dl} that even for $x_m/L=1$ the above analytical expression agrees very well with the numerical results over the entire range of $\Delta L^c$. For higher values of $x_m/L$, Eq.\ (\ref{eq:dist_dlc}) gives a reasonable description of the distribution for small $\Delta L^c$ values. It  is important to note that for large upper bounds $x_m$, in the regime of large misalignment $x\gg L$ the critical elongation becomes a linear function of $x$, hence, the distribution $p(\Delta L^c)$ can be well approximated by a uniform distribution. This behaviour can be observed in Fig.\ \ref{fig:dist_dl} for $x_m/L=100$.

A similar analysis can be carried out for the force $F_y$ kept by the fibers. Starting from Eq.\ (\ref{eq:fyfiber}) it can shown that in the limiting case of $x_m/L\ll 1$ and $\Delta L \ll L$ the fibers keep practically the same load, the dependence on the misalignment is negligible. It follow that under the above conditions the behaviour of our FBM where fibers have the same strength and the only source of disorder is the misalignment of fibers, can be approximated as an equal load sharing FBM of perfectly aligned fibers, where fibers have a random strength described by the probability density function Eq.\ (\ref{eq:dist_dlc}). The macroscopic stress-strain curve $\sigma(\varepsilon)$ of ELS FBMs can be obtained analytically as
\begin{align}
    \sigma = E\varepsilon\left[1-P(\varepsilon)\right],
    \label{eq:elsconstit}
\end{align}
where $P(x)$ denotes the cumulative distribution of fibers' strength \cite{hansen2015fiber}. The expression $E\varepsilon$ is the load kept by a single intact fiber, while the term $\left[1-P(E\varepsilon)\right]$ yields the fraction of intact fibers at the strain $\varepsilon$.
\begin{figure}
\begin{center}
\epsfig{file=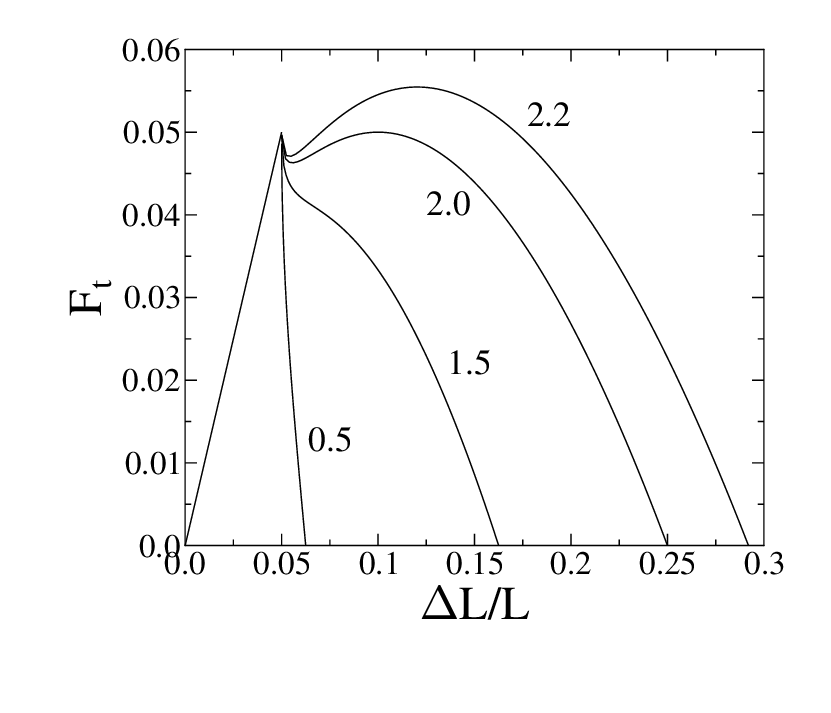,bbllx=40,bblly=20,bburx=370,bbury=330, width=8.5cm}
\caption{Constitutive behaviour of a fiber bundle where fibers are perfectly aligned ($x=0$ for all fibers) but have a random strength sampled from the distribution Eq.\ (\ref{eq:dist_dlc}). The numbers placed next to the curves indicate the value of $x_m/L$.
}   
\label{fig:analit_constit}
\end{center}
\end{figure} 
The cumulative distribution $P(\Delta L^c)$ of the threshold elongations can be obtained from the probability density function Eq.\ (\ref{eq:dist_dlc}) as
\begin{align}
    P(\Delta L^c) = \frac{L}{x_m} \left(\frac{\Delta L^c}{L\varepsilon_{th}}-1\right)^{1/2}.
    \label{eq:map_cumul}
\end{align}
Substituting this form into the general expression Eq.\ (\ref{eq:elsconstit}), the constitutive equation of our ELS FBM reads as
\begin{align}
    F_t(\Delta L) = \Delta L\left[1-\frac{L}{x_m}\left(\frac{\Delta L}{L\varepsilon_{th}}-1\right)^{1/2}\right], 
    \label{eq:constit_elsfbm}
\end{align}
where $\Delta L$ goes from $L\varepsilon_{th}$ to $L\varepsilon_{th}(1+x_m^2/L^2)$. Between 0 and $L\varepsilon_{th}$ the force $F_t$ has a linear dependence on $\Delta L$.
Figure \ref{fig:analit_constit} illustrates the behaviour of the analytical expression Eq.\ (\ref{eq:constit_elsfbm}) for several values of $x_m$. A strong qualitative similarity is obtained between the evolution of $F_t(\Delta L)$ of Eq.\ (\ref{eq:constit_elsfbm}) with the degree of disorder $x_m$ and the behaviour of the FBM with misaligned fibers presented in Fig.\ \ref{fig:constit}. The results demonstrate that the behaviour of our FBM with structural disorder but no strength disorder can be mapped to an ELS FBM where fibers are perfectly aligned and the strength of fibers is the only stochastic variable. 

\section{Finite size scaling}
The analytical expression of the constitutive equation Eq.\ (\ref{eq:constit}) of the misaligned fiber bundle corresponds to the infinite system size. However, for finite bundles considered in the numerical calculations, characteristic quantities of the system have a non-trivial dependence on the number of fibers $N$. To quantify this size effect we carried out computer simulations to study how the strength of the bundle and the avalanche size distribution evolve as the number of fibers $N$ is varied from 500 to 500,000.
\begin{figure}
\begin{center}
\epsfig{file=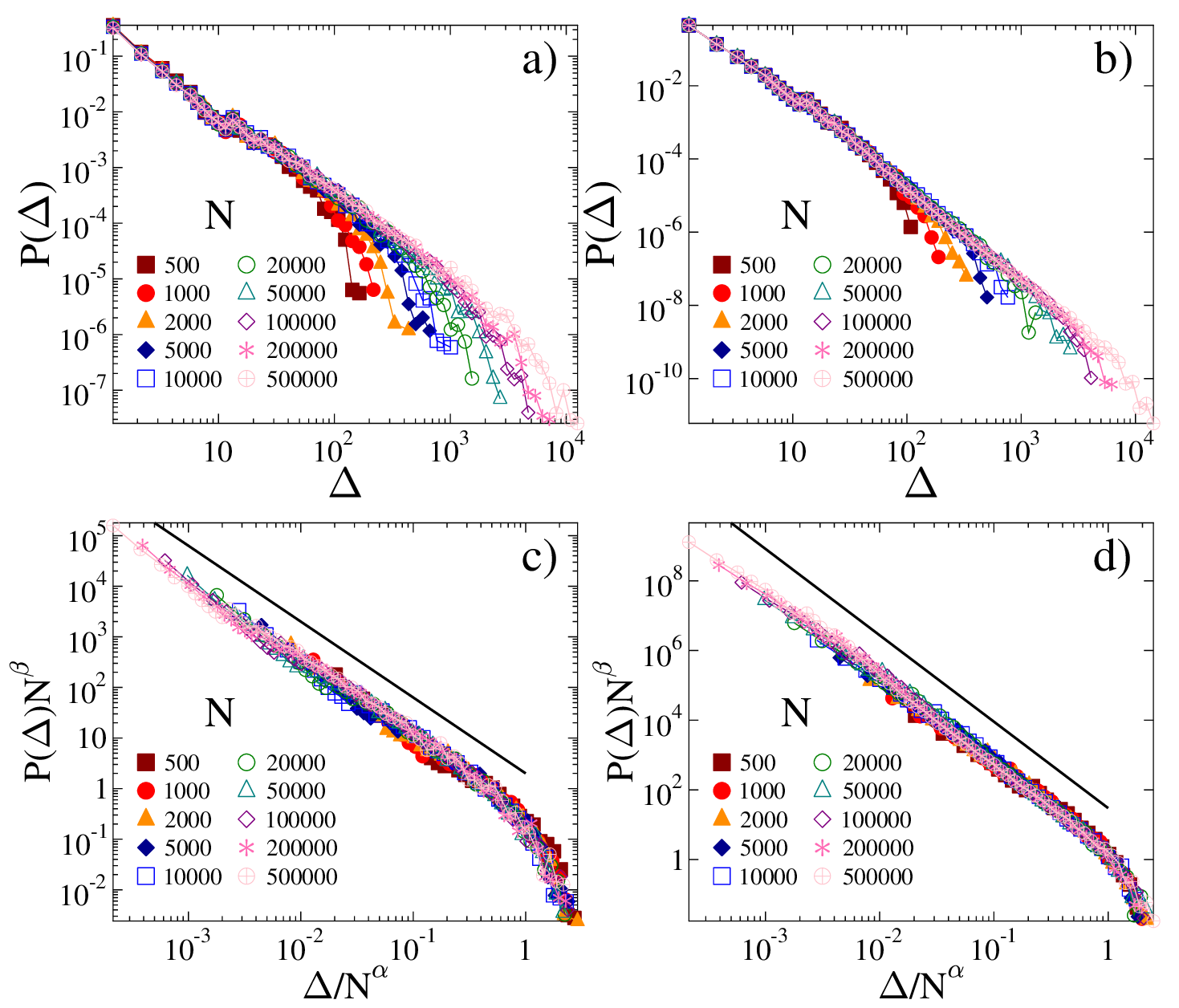,bbllx=20,bblly=0,bburx=695,bbury=590, width=8.5cm}
\caption{Avalanche size distributions $p(\Delta)$ for several system sizes $N$ at the degree of misalignment $x_m=4.14$ $(a)$ and $x_m=10$ $(b)$. Rescaling the data of $(a)$ and $(b)$ with appropriate powers $\alpha$ and $\beta$ of the number of fibers $N$ along the horizontal and vertical axis in $(c)$ and $(d)$, the curves obtained at different system sizes can be collapsed on the top of each other. The value of the scaling exponents are $\alpha=2/3$, $\beta=1$ $(c)$, and $\alpha=2/3$, $\beta=5/3$ $(d)$. The straight lines represent power laws of exponent $\tau = 3/2$ $(c)$ and $\tau=5/2$ $(d)$.
}
\label{fig:aval_size}
\end{center}
\end{figure}

For clarity, Figures \ref{fig:aval_size}$(a,b)$ present the size distribution of avalanches $p(\Delta)$ obtained at different system sizes $N$ for two values of the degree of structural disorder $x_m$, i.e.\ close to the critical disorder (Fig.\ \ref{fig:aval_size}$(a)$) and far from it (Fig.\ \ref{fig:aval_size}$(b)$), where single power laws are expected with exponents $\tau=3/2$ and $\tau=5/2$, respectively. The cutoff avalanche size of the distributions increases with the number of fibers $N$, which indicates that in larger systems larger avalanches can be generated, however, the value of the exponent $\tau$ remains the same. Figure \ref{fig:aval_size}$(c,d)$ demonstrate that rescaling the data of Figs. \ref{fig:aval_size}$(a,b)$ with appropriate powers $\alpha$ and $\beta$ of $N$ along the horizontal and vertical axis the avalanche size distributions of different system sizes can be collapsed on a master curve. The value of $\alpha$ providing best collapse is the same for both $x_m$ values $\alpha=2/3$, however, $\beta$ has different values $\beta=1$ and $\beta=5/3$ in Fig.\ \ref{fig:aval_size}$(c)$ and Fig.\ \ref{fig:aval_size}$(d)$, respectively. The good quality data collapse implies that avalanche size distributions $p(\Delta, N)$ measured at different system sizes obey the scaling structure
\begin{align}
p(\Delta, N) = N^{-\beta}\Phi\left(\frac{\Delta}{N^{\alpha}}\right),
\end{align}
where the scaling function $\Phi(x)$ has a power law behaviour $\Phi(x)\sim x^{-\tau}$. It is important to emphasize that the exponents fullfill the scaling law
\begin{align}
 \beta = \tau \alpha
\end{align}
with a high accuracy.

\begin{figure}
\begin{center}
\epsfig{file=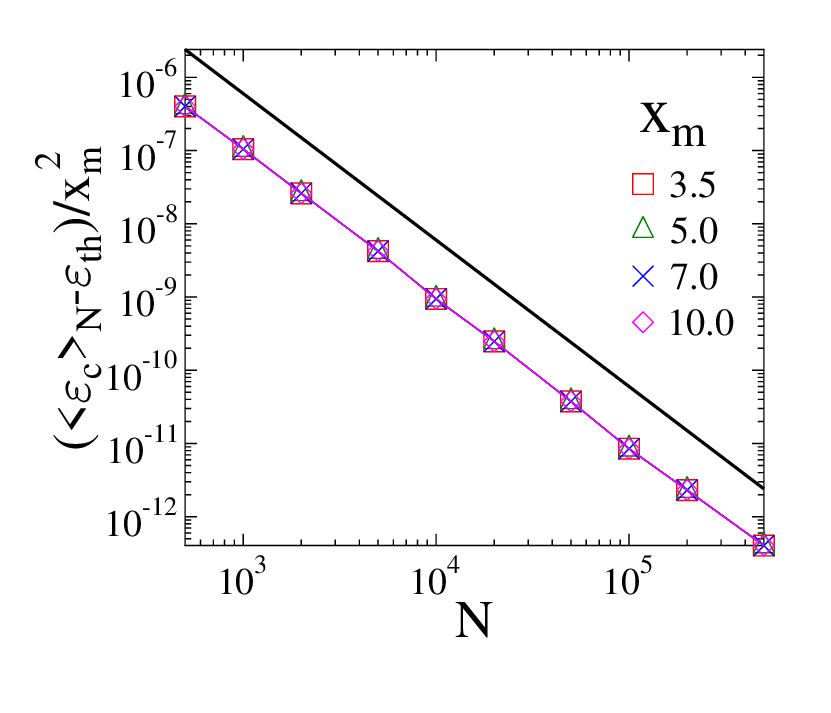,bbllx=20,bblly=20,bburx=380,bbury=330, width=8.5cm}
\caption{
Average value of the critical strain $\left<\varepsilon_c\right>_N$ in the brittle phase of the system as a function of the number of fibers $N$ for four values of the degree of disorder $x_m$. From each data set we subtracted the asymptotic strength $\varepsilon_{th}$. The straight line represents a decreasing power law of exponent $2$. Rescaling the data along the vertical axis by $x_m^2$, the curves of different $x_m$ values collapse on the top of each other.}
\label{fig:scaling_dl}
\end{center}
\end{figure}

In the brittle phase of the bundle $x_m<x_c$, the first beam breaking triggers a catastrophic avalance, hence, the failure strain $\varepsilon_c$ is determined by the smallest failure threshold of fibers as $\varepsilon_c=\Delta L^c_{min}/L$. In a bundle of $N$ fibers the average value of the smallest failure threshold $\left<\Delta L^c_{min}\right>_N$ can be obtained starting from the cummulative distribution Eq.\ (\ref{eq:map_cumul})
\begin{align}
\left<\Delta L^c_{min}\right>_N = P^{-1}\left(\frac{1}{N+1}\right),
\end{align}
where $P^{-1}$ denotes the inverse of the distribution function \cite{hansen_statistical_2000}.
Substituting Eq.\ (\ref{eq:map_cumul}), the size dependence of the critical elongation at the brittle peak can be cast into the form
\begin{align}
 \left<\Delta L^c_{min}\right>_N = L\varepsilon_{th} + \frac{x_m^2\varepsilon_{th}}{L}N^{-2}.
 \label{eq:brit_sizescaling}
\end{align}
The result indicates that as the size of the bundle increases the value of the critical strain $\left<\varepsilon_c\right>_N=\left<\Delta L^c_{min}\right>_N/L$ decreases towards the asymptotic limit $\left<\varepsilon_c\right>_N \to \varepsilon_{th}$ according to a power law of exponent 2. Figure \ref{fig:scaling_dl} demonstrates that the numerical results of size scaling have a perfect agreement with Eq.\ (\ref{eq:brit_sizescaling}). The analytical expression also shows that the degree of disorder $x_m$ affects the convergence to the asymptotic value. In Figure \ref{fig:scaling_dl} the data corresponding to different $x_m$ values are scaled with $x_m^2$ which resulted in an excellent collapse of the curves, in agreement with Eq.\ (\ref{eq:brit_sizescaling}).

\begin{figure}
\begin{center}
\epsfig{file=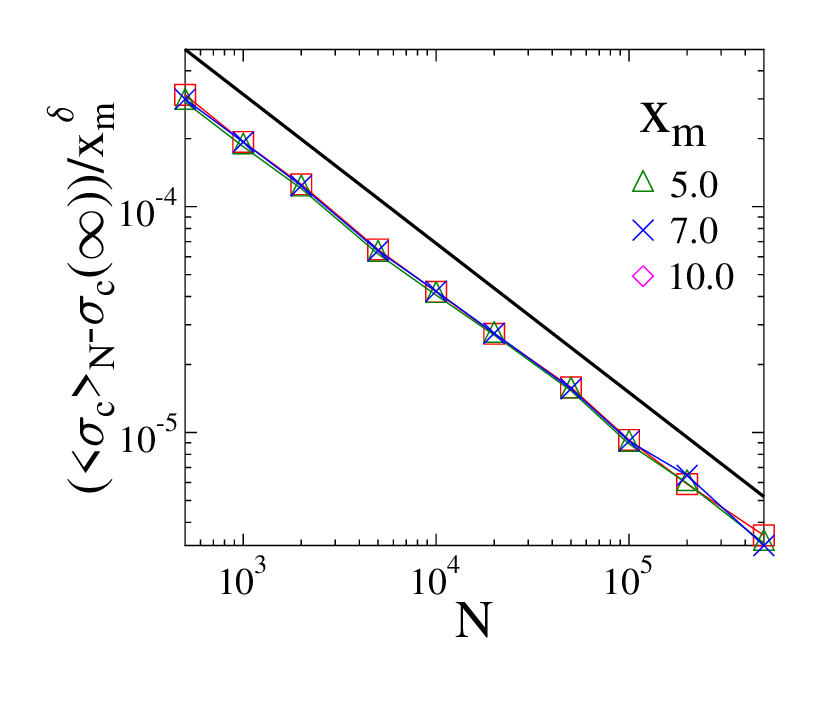,bbllx=20,bblly=20,bburx=380,bbury=330, width=8.5cm}
\caption{Average value of the critical stress $\left<\sigma_c\right>_N$ in the quasi-brittle phase of the system as a function of the number of fibers $N$ for three values of the degree of disorder $x_m$. The straight line represents the power law of exponent $\alpha=2/3$. The value of $\delta$ providing best collapse is $\delta=2/3$.
}
\label{fig:scaling_fc}
\end{center}
\end{figure}

As a second example for the size scaling of macroscopic characteristics of the misaligned fiber bundle, we analyzed how the critical stress $\sigma_c=F_c/N$ depends on the number of fibers $N$ in the quasi brittle regime $x_m>x_c$. Figure \ref{fig:scaling_fc} demonstrates that with increasing $N$ the curves of $\left<\sigma_c\right>_N$ converge to finite asymptotic values $\sigma_c(\infty)$ at each disorder $x_m$. Subtracting the proper value $\sigma_c(\infty)$, power laws are obtained which are again collapsed on the top of each other by rescaling with an appropriate power $\delta$ of the degree of disorder $x_m$. The numerical results imply the functional form
\begin{align}
\left<\sigma_c\right>_N = \sigma_c(\infty) + AN^{-\alpha}x_m^{\delta},
\end{align}
where $A$ is a multiplication factor.

The scaling analysis presented in Fig.\ \ref{fig:scaling_fc} revealed that the exponent $\alpha$ has the same value $\alpha=2/3$ as for the cutoff of the avalanche size distributions in Figs.\ \ref{fig:aval_size}$(c, d)$. The value of the exponent $\delta$, which describes the effect of disorder on the size scaling of fracture strength, proved to be $\delta=2/3$ providing best collapse in the figure. The asymptotic values $\sigma_c(\infty)$ can be obtained from the analytic solution of the constitutive curve Eq.\ (\ref{eq:constit}).

\section{Discussion}
Disorder is an inherent property  of natural materials and also of many of the artificially made ones. Understanding the role of disorder in fracture processes has a great importance with a lot of interesting challenges both for theoretical studies and practical applications. Among theoretical approaches, the fiber bundle model is able to capture the relevant mechanisms of the fracture of disordered materials but at the same time its simplicity makes it possible to cast several characteristic quantities into analytical forms as a function of the parameters of the system.

In real fibrous structures fiber orientation may not be completely aligned so that not all fibers can be parallel to the load axis. This misalignment of fibers can affect both the macroscopic response of fibrous system and the microscopic process of their fracture under an increasing load. To investigate this effect, here we considered a fiber bundle model which has a disordered micro-structure, i.e.\ fibers are allowed to be misaligned in such a way that the fibers' two end points are displaced with a random amount in the direction perpendicular to the bundle axis. We analyzed the behaviour of the model focusing on the effect of the degree of structural disorder which is controlled through a parameter of the probability distribution of the misalignment of fibers. To isolate the effect of structural disorder, in our approach no strength disorder is considered, i.e.\ fibers are assumed to break at the same local strain.

Misalignment has the consequence that even if the bundle is stretched between two hard plates, the load sharing after fiber breaking is not equal: the excess load, intact fibers receive from a broken one, depends on the fibers' misalignment, which implies a global but not equal load sharing. To analyze the macroscopic response of the bundle we could cast the force-elongation equation as an integral over the stochastic misalignment. We demonstrated analytically and numerically that for small elongations the behaviour of the misaligned fiber bundle is linear, however, both the stiffness and fracture strength of the bundle decrease with increasing misalignment. For low misalignment the end point of the initial linear regime is a global maximum of the force-elongation curve, which implies that the breaking of the first fiber triggers the immediate abrupt failure of the bundle. In this perfectly brittle phase the bundle cannot tolerate any damage, however, as the amount of structural disorder increases, the force-elongation curve develops a second maximum, which becomes gradually higher than the brittle one. Above the critical disorder the response of the bundle becomes quasi-brittle, where global failure is approached through stable avalanches of breaking events. Further increasing the structural disorder the ultimate strength of the bundle increases while the amount of damage the system can sustain saturates to a limit value. 

The disorder driven brittle-quasi-brittle transition has been widely studied before varying the strength disorder of fibers in the absence of structural randomness \cite{ray_epl_2015,roy_phasetrans_pre2019,hidalgo_universality_2008,PhysRevE.87.042816,roy_prres_2019,menezes-sobrinho_influence_2010,kim_phase_2004,danku_disorder_2016}. It was found that in the vicinity of critical disorder characteristic quantities of the system exhibit scaling, and the transition occurs analogous to continuous phase transition. Here we demonstrated that structural disorder alone makes the transition abrupt characterized by a finite jump of the critical elongation and damage at the brittle-quasi-brittle  critical point, and by the absence of scaling. We established a mapping between our FBM of structural disorder and heterogeneous load distribution to an ELS FBM where fibers are perfectly aligned but have strength disorder. The mapping provides an adequate description of the behaviour of misaligned bundles for sufficiently low disorder and low deformations. Based on the mapping we could demonstrate that first order type brittle - quasi-brittle transition can also occur in ELS FBMs with properly selected strength disorder.

In the calculations we used uniformly distributed misalignment values between 0 and an upper bound, which had the advantage that, on the one hand, it allowed for analytical calculations, and on the other hand, we could easily control the degree of disorder by varying the upper bound of misalignment values. However, it is important to emphasize that the qualitative behaviour of the fiber bundle does not change if misalignment values are distributed over an infinite interval. For instance, in the case of the exponential distribution $p(x)=(1/\lambda)\exp{\left(-x/\lambda\right)}$ between 0 and $+\infty$, at low values of the average misalignment $\lambda$, brittle failure occurs at the fixed threshold strain $\varepsilon_{th}$ due to the high fraction of fibers nearly aligned with the bundle axis. Gradually increasing $\lambda$, the force-elongation curve develops a second maximum, which becomes higher than the brittle peak at a critical value $\lambda_c$.

From the viewpoint of applications, a very important consequence of our results is that an aligned fiber bundle which has a brittle behaviour due to the nearly constant strength of fibers, can be stabilized by randomizing its structure. Introducing a sufficient degree of structural disorder in the form of misalignment of fibers can make the failure process quasi-brittle, where macroscopic failure is preceded by breaking avalanches, and the fracture strength of the bundle exceeds the one of its aligned counterpart.

It is also a very interesting feature of our system that in spite of the inhomogeneous distribution of load on fibers, breaking avalanches have a power law behaviour. As the amount of structural disorder increases a crossover emerges between two power law regimes of different exponents. The value of the exponents is equal to the mean field avalanche exponents of FBMs which have been found before analytically \cite{pradhan_crossover_2005-1,pradhan_crossover_2006,hansen2015fiber}. Our simulations showed that both the macroscopic strength and the size distribution of avalanches depend on the size of the bundle. We performed a scaling analysis which revealed that in larger systems larger breaking avalanches can emerge, and the macroscopic strength parameters converge towards finite asymptotic values, all described by scaling exponents. Our study demonstrates that structural disorder results in a highly complex behaviour of the failure process of fibres structures. Of course, when two sources of disorder are present in the system, i.e.\ structural and strength disorder,further interesting effects are expected which will be explored in a forthcoming publication.

\begin{acknowledgments}
Supported by the University of Debrecen Program for Scientific Publication. Project no. TKP2021-NKTA-34 has been implemented with the support provided from the National Research, Development and Innovation Fund of Hungary, financed under the TKP2021-NKTA funding scheme.
\end{acknowledgments}

\bibliography{statphys_fracture}

\end{document}